\documentclass[aps,prd,doublecol,showpacs,showkeys,amsmath,amssymb]{epl2}
\usepackage{amsmath,amsfonts,graphicx,psfrag,subfigure,fancyhdr}
\usepackage[small,nooneline]{caption2}

\title{Octet Quark Contents from SU(3) Flavor Symmetry}

\author{Haitao Liu\inst{1}, Yujie Chi\inst{1}, Lijing Shao\inst{1} \and Bo-Qiang Ma\inst{1,2}}

\institute{
  \inst{1} School of Physics and State Key Laboratory of Nuclear Physics and Technology, Peking University, Beijing 100871, China\\
  \inst{2} Center for High Energy Physics, Peking University, Beijing 100871, China
} %


\abstract{With the parametrization of parton distribution functions
(PDFs) of the proton by Soffer \textit{et al.}, we extend the
valence quark contents to other octet baryons by utilizing SU(3)
flavor symmetry. We find the method practically useful.
Fragmentation functions (FFs) are further obtained through the
phenomenological Gribov-Lipatov relation at the $x \to 1$ region.
Our results are compared with different models, and these different
predictions can be discriminated by upcoming experiments.}

\begin{document}

\maketitle

\section{Introduction}\label{secint}

A major property of the theory of strong interactions, \textit{i.e.}
quantum chromodynamics (QCD), is color confinement, i.e., quarks and
gluons cannot be free but must be bounded inside hadrons. Due to the
complicated non-perturbative effects of strong interaction,
nowadays, it is still difficult to calculate baryon structure
functions, in terms of parton distribution functions (PDFs), from
the basic principle of QCD. One possible direction is lattice QCD,
however, its calculation is largely constrained by the computational
capability. Therefore, phenomenological models based on the spirit
of QCD still play an important role in the studies of baryon
structures.

Statistical models, providing intuitive appeal and physical
simplicity, have made amazing success. Angelini and Pazzi first
verified that valence quark distribution has a statistical behavior
when the Bjorken variable $x$ is larger than $0.1$. Utilizing a
thermodynamical form as an input to the first order QCD
$Q^{2}$-evolution, they obtained a good fit at
$5<Q^{2}<100~{\textrm{GeV}^2}$~\cite{angelini1angelini2angelini3}.
Cleymans \textit{et al.} then argued that Pauli principle should
have effects on PDFs, and based on the MIT-bag model they got a
statistical way to generate reasonable
PDFs~\cite{cleyman1cleyman2cleyman3}. Along this way, several other
models were proposed~\cite{mac,bickerstaff,devanathan,dey, bhalerao,
trevisan,zhangyh1zhangyh2,bbs,bbs2,bbs3}, where quarks follow
Fermi-Dirac distribution and gluons obey Bose-Einstein statistics.
For example, Mac and Ugaz~\cite{mac} incorporated first order QCD
correction and Devanathan \textit{et al.}~\cite{devanathan}
exhibited the scaling behavior. Afterwards, Bhalerao \textit{et
al.}~\cite{bhalerao} considered finite-size correction, and Trevisan
\textit{et al.}~\cite{trevisan} introduced a confining potential and
took gluon splitting into account. Recently, Zhang \textit{et al.}
established a new and simple statistical model in terms of
light-front kinematic variables and calculated the nucleon
PDFs~\cite{zhangyh1zhangyh2}. Besides, models based on other
statistical principles were also studied. For instance, Zhang
\textit{et al.}~\cite{zhang1zhang2zhang3} constructed a model using
the principle of detailed balance, and the Gottfried sum they got is
in surprising agreement with experiments. Singh and
Upadhyay~\cite{singh} extended this model to have spin considered
with modifications. Further, Alberg and Henley~\cite{alberg} tracked
the detailed balance model, and obtained PDFs for the proton as well
as the pion.

BBS model, established by Bourrely, Buccella and
Soffer~\cite{bbs,bbs2,bbs3}, allows PDFs to be directly expressed in
terms of $x$. By using 8 free parameters, the model can obtain all
polarized and unpolarized PDFs of the proton, and it well fits to
experimental data in a broad range of $x$ and $Q^{2}$. Extension of
the model to other baryons is still in progress, but one of the
limitations is in the determination of unknown parameters due to the
currently relatively lack of relevant experiments. We stress that,
SU(3) flavor symmetry~\cite{su3} presents a kind of connection of
quark distributions between octet baryons, and by using it, we get
PDFs of all octet baryons from the BBS PDFs of the proton.

On the other hand, fragmentation function (FF), which describes
the process of quark hadronization, is another basic quantity in QCD.
One application of PDFs is to study the fragmentation function
$D_{q}(z)$. At present, it will be very useful if there exists
simple connections between them, so that one can predict the poorly
known $D_{q}(z)$ from $q(x)$. One tentative attempt is the so-called
Gribov-Lipatov relation~\cite{glrelation}, where
\begin{equation}
x q^{h}(x)= D^{h}_{q}(z),
\end{equation}
is suggested as a phenomenological approximate when $z (=x) \to 1$.
Utilizing this relation, we get FF ratios, from our calculated PDF
ratios, $u(x)/d(x)$ and $u(x)/s(x)$, for $p$ and $\Lambda$, at the
$z\to 1$ region.  Recently, Soffer \textit{et al.} gave FFs out of
the same statistical spirit, and they parameterized the free
parameters for $p$ and $\Lambda$ directly from experimental
data~\cite{ff2003}. On the contrary, our results are based on SU(3)
flavor symmetry.

The paper is organized as follows. First we briefly review the
quark-diquark model, perturbative QCD (pQCD), and BBS model. Then we
extend the BBS PDFs to octet baryons through SU(3) flavor symmetry,
and compare results among different models.

\section{Quark-diquark model}\label{secqaq}

SU(6) quark-diquark model, based on the SU(6) quark wave-functions,
was applied to nucleon structures when $x$ is large, first by
Feynman~\cite{feynman}. In this model, when a valence quark of
baryon is probed, we can reorganize the other two quarks as a
mixture of a scalar diquark with spin 0 and an axial vector diquark
with spin 1, denoted as $S$ and $V$ respectively.

Therefore, the unpolarized PDF is written as
\begin{equation}
q(x)=c_{q}^{\scriptscriptstyle{S}}a_{\scriptscriptstyle{S}}(x)+c_{q}^{\scriptscriptstyle{V}}a_{\scriptscriptstyle{V}}(x),
\end{equation}
where $c_{q}^{\scriptscriptstyle{S}}$ and
$c_{q}^{\scriptscriptstyle{V}}$, the weight coefficients, can
be determined from SU(6) quark-diquark wave-functions, and their
values are different for different baryons;
$a_{\scriptscriptstyle{D}}(x)$ ($D=S,V$) represents the probability
of finding the diquark in the state $D$, and when expressed in terms of
light-cone wave-function~\cite{22}, is
\begin{equation}
a_{\scriptscriptstyle{D}}(x)\propto\int[d^{2}\vec{k}_{\bot}]|\varphi(x,\vec{k}_{\bot})|^{2},
\end{equation}
with normalization to 3 when integrated in $x \in [0,1]$.

To calculate explicitly, we can use the Brodsky-Huang-Lepage (BHL)
prescription for the light-cone wave-function in momentum space,
\begin{equation}
\varphi(x,\vec{k}_{\bot}) \propto
\exp\left\{-\frac{1}{8\alpha_{\scriptscriptstyle{D}}^{2}}
[\frac{m_{q}^{2}+\vec{k}_{\bot}^{2}}{x}+\frac{m_{\scriptscriptstyle{D}}^{2}+\vec{k}_{\bot}^{2}}{1-x}]\right\},
\end{equation}
where $m_{q}$ and $m_{\scriptscriptstyle{D}}$ are masses of quarks
and diquarks, respectively, and the parameter
$\alpha_{\scriptscriptstyle{D}} \simeq 330~{\rm MeV}$.

Similarly, for a polarized baryon, the PDF reads
\begin{equation}
{\Delta}q(x)=\tilde{c}_{q}^{\scriptscriptstyle{S}}\tilde{a}_{\scriptscriptstyle{S}}(x)+
\tilde{c}_{q}^{\scriptscriptstyle{V}}\tilde{a}_{\scriptscriptstyle{V}}(x),
\end{equation}
\begin{equation}
\tilde{a}_{\scriptscriptstyle{D}}(x) =
\int[d^{2}\vec{k}_{\bot}]|\varphi(x,\vec{k}_{\bot})|^{2}|W_{\scriptscriptstyle{D}}(x,\vec{k}_{\bot})|,
\end{equation}
\begin{equation}
W_{\scriptscriptstyle{D}}(x,k_{\bot})=\frac{(k^{+}+m_{q})^2-\vec{k}_{\bot}^{2}}{(k^{+}+m_{q})^{2}+\vec{k}_{\bot}^{2}},
\end{equation}
where
$k^{+}=x\sqrt{({m_{q}^{2}+\vec{k}_{\bot}^{2}})/{x}+({m_{\scriptscriptstyle{D}}^{2}+\vec{k}_{\bot}^{2}})/({1-x})}$.

The PDFs of five baryons calculated from the quark-diquark model are
listed in Table~\ref{1}~\cite{23}.

\section{Perturbative QCD }\label{secpqcd}

In pQCD, the quark distribution in the hadron $h$ satisfies the
counting rule~\cite{24},
\begin{equation}
q_{h}(x)\propto (1-x)^{p}, \quad p=2n-1+2{\Delta}S_{z},
\end{equation}
where $n$ is the minimal number of spectator quarks, and
${\Delta}S_{z} = |S^{q}_{z}-S^{h}_{z}|$ equals $0$ and $1$ for
parallel and anti-parallel quark helicity with respect to the
hadron~\cite{25}.

Keeping to the next-to-leading order,
\begin{equation}
q_{i}^{\uparrow}(x)=\frac{\tilde{A}_{q_{i}}}{B_{3}}x^{-1/2}(1-x)^{3}+\frac{\tilde{B}_{q_{i}}}{B_{4}}x^{-1/2}(1-x)^{4},
\end{equation}
\begin{equation}
q_{i}^{\downarrow}(x)=\frac{\tilde{C}_{q_{i}}}{B_{5}}x^{-1/2}(1-x)^{5}+\frac{\tilde{D}_{q_{i}}}{B_{6}}x^{-1/2}(1-x)^{6},
\end{equation}
where $B_{n}=B(1/2,n+1)$ is $\beta$-function. For each baryon, there
are 5 constraints among 8 parameters~\cite{23}, so only 3 are left
free. Parameters for all octet baryons are given in Table~\ref{2}.

A more brief form of pQCD (called brief pQCD hereafter),
considering only the leading terms of quark helicity distribution, is
expressed as~\cite{26}
\begin{equation}\label{pqcd2a}
q_{i}^{\uparrow}(x)=\frac{\tilde{A}_{q_{i}}}{B_{3}}x^{-1/2}(1-x)^{3},
\end{equation}
\begin{equation}\label{pqcd2b}
q_{i}^{\downarrow}(x)=\frac{\tilde{C}_{q_{i}}}{B_{5}}x^{-1/2}(1-x)^{5}.
\end{equation}
For the nucleon, 4 constraints are given. They are (1) normalizations
for $q_{i}$, \textsl{i.e.,} $N_{i}=\tilde{A}_{q_{i}}+\tilde{C}_{q_{i}}$; and (2) the
corresponding polarizations, \textsl{i.e.,}
${\Delta}q_{i}=\tilde{A}_{q_{i}}-\tilde{C}_{q_{i}}$. Parameters can
be extracted through
$\Sigma={\Delta}u+{\Delta}d+{\Delta}s\approx0.3$, and the Bjorken
sum rule $\Gamma^{p}-\Gamma^{n}=\frac{1}{6}g_{A}/g_{V}\approx0.2$,
from polarized DIS experiments~\cite{27}. Besides, all octet baryons
are related to each other by SU(3) flavor symmetry. So there is no
free parameter in the brief pQCD.  All parameters are listed in
Table~\ref{pqcd2}.

\section{BBS model}\label{secbbs}

In the BBS statistical approach, the nucleon is assumed to be a
massless parton gas in equilibrium at a temperature $T$. The parton
distribution $p(x)$, at an input energy scale $Q_{0}^{2}$, can be
written as
\begin{equation}
p(x,Q_{0}^{2})\propto[e^{(x-X_{0p})/\bar{x}}\pm1]^{-1},
\end{equation}
where ``$+$'' is for fermions (quarks and anti\mbox{}quarks) and
``$-$''for bosons (gluons). $X_{0p}$ is the thermodynamical
potential of parton $p$, and $\bar{x}$ is the universal temperature.

PDFs for $u$, $d$ quarks of helicity $h$, are~\cite{bbs}
\begin{equation}
xq^{h}(x,Q_{0}^{2})=\frac{AX^{h}_{0q}x^{b}}{e^{(x-X^{h}_{0q})/\bar{x}}+1}+\frac{\tilde{A}x^{\bar{b}}}{e^{x/\bar{x}}+1},
\end{equation}
\begin{equation}
x\bar{q}^{h}(x,Q_{0}^{2})=\frac{\bar{A}(X^{-h
}_{0q})^{-1}x^{2b}}{e^{(x+X^{-h}_{0q})/\bar{x}}+1}+\frac{\tilde{A}x^{\bar{b}}}{e^{x/\bar{x}}+1},
\end{equation}
where we have
$q_{i}=q_{i}^{+}+q_{i}^{-},~~{\Delta}q_{i}=q_{i}^{+}-q_{i}^{-}$.
From a next-to-leading-order fit of a selection of 233 data points
at $Q_{0}^{2}=4~{\mathrm{GeV}}^{2}$, parameters for the proton were
determined,
\begin{eqnarray}
\bar{x}=0.099,  \quad b=0.40962,  \quad \tilde{b}=-0.25347, \\ \nonumber %
 A=1.74938, \quad \tilde{A}=0.08318,   \quad \bar{A}=1.90801, \\ \nonumber %
X_{od}^{-}=0.30174, \quad X_{od}^{+}=0.22775,  \\ \nonumber %
X_{ou}^{+}=0.46128, \quad X_{ou}^{-}=0.29766.
\end{eqnarray}

Considering the asymmetry, PDFs for $s$, $\overline{s}$
quarks~\cite{bbs3} are formulated as
\begin{equation}
\begin{split}
xs^{h}(x,Q_{0}^{2}) =&
\frac{AX^{+}_{0u}x^{b_{s}}}{e^{\left(x-X^{h}_{0s}\right)/\bar{x}}+1}\frac{\ln\left(1+e^{kX_{0s}^{h}/\bar{x}}\right)}{\ln\left(1+e^{kX_{0u}^{+}/\bar{x}}\right)}
\\ &+\frac{\tilde{A}_{s}x^{\tilde{b}}}{e^{x/\bar{x}}+1},
\end{split}
\end{equation}
\begin{equation}
\begin{split}
x\bar{s}^{h}(x,Q_{0}^{2})
={}&\frac{\bar{A}\left(X^{+}_{0d}\right)^{-1}x^{2b_{s}}}{e^{\left(x+X^{-h}_{0s}\right)/\bar{x}}+1}
\frac{\ln\left(1+e^{-kX_{0s}^{-h}/\bar{x}}\right)}{\ln\left(1+e^{-kX_{0d}^{+}/\bar{x}}\right)}{}\\
&+\frac{\tilde{A}_{s}x^{\tilde{b}}}{e^{x/\bar{x}}+1}.
\end{split}
\end{equation}
For the proton, the parameters are,
\begin{eqnarray}
X_{0s}^{+}=0.08101, \quad X_{0s}^{-}=0.2029, \\ \nonumber %
b_{s}=2.05305, \quad \tilde{A}_{s}=0.05762.
\end{eqnarray}

From above expressions and by utilizing SU(3) flavor symmetry (Table ~\ref{symmetry}), we
get PDFs for all octet baryons as shown in
Fig.~1 to Fig.~5.

\section{Fragmentation functions}\label{secfrag}

Fragmentation function for $u$,~$d$,~$s$ quarks, given by Soffer and
Bourrely~\cite{ff2003}, is of the form,
\begin{equation}
D^{\scriptscriptstyle{B}}_{q}(x,Q^{2}_{0})=\frac{A^{\scriptscriptstyle{B}}_{q}X^{\scriptscriptstyle{\scriptscriptstyle{B}}}_{q}x^{b}}{e^{\left(x-X^{\scriptscriptstyle{B}}_{q}\right)/\bar{x}}+1},
\end{equation}
where $X^{\scriptscriptstyle{B}}_{q}$ is the potential related to
the fragmentation process $q \to B$. They then determine the parameters 
as~\cite{ff2003}
\begin{eqnarray}
A^{\Lambda}_{u}=A^{\Lambda}_{d}=0.428, \quad b=0.200, \quad
\bar{x}=0.099,
\\ \nonumber %
A^{p}_{u}=A^{p}_{d}=0.264, \quad A^{p}_{s}=1.168, \quad
A^{\Lambda}_{s}=1.094,
\\ \nonumber
X^{p}_{u}=0.648, \quad X^{p}_{s}=0.247, \\ \nonumber %
X^{\Lambda}_{u}=0.296, \quad X^{\Lambda}_{s}=0.476,
\end{eqnarray}
where the input energy scale was fixed at
$Q_{0}=0.632~{\mathrm{GeV}}$, and the yielded $\chi^{2} = 227.5$ for
206 experimental points.

\section{Comparisons between models}\label{secdis}

We calculate PDFs as a function of $x$ for models mentioned above,
and our extension of BBS model based on SU(3) flavor symmetry is
compared with them. Comparisons are shown in Figs.~$1,2,3,4,5$.

Below we discuss physical issues at the $x \to 1$ region.

In Fig.~1, we illustrate
${\Delta}u_{\scriptscriptstyle{V}}(x)/u_{\scriptscriptstyle{V}}(x)$
of the proton. We can see that quark-diquark model, pQCD model, and
brief pQCD model all give
${\Delta}u_{\scriptscriptstyle{V}}(x)/u_{\scriptscriptstyle{V}}(x)\rightarrow1$
when $x \to 1$, that is, the proton is completely positively
polarized. However, BBS model gives a different result of 0.779.
Fig.~1 also shows
${\Delta}d_{\scriptscriptstyle{V}}(x)/d_{\scriptscriptstyle{V}}(x),~d_{\scriptscriptstyle{V}}(x)/u_{\scriptscriptstyle{V}}(x)$
for the proton. For
${\Delta}d_{\scriptscriptstyle{V}}(x)/d_{\scriptscriptstyle{V}}(x)$,
pQCD model and brief pQCD model give
${\Delta}d_{\scriptscriptstyle{V}}(x)/d_{\scriptscriptstyle{V}}(x)
\rightarrow 1$ when $x \to 1$, while quark-diquark model and BBS
model predict negative values. For
$d_{\scriptscriptstyle{V}}(x)/u_{\scriptscriptstyle{V}}(x)$, the
prediction of BBS model is close to those from pQCD and brief pQCD
models, while the result of quark-diquark model is different from
that of other models, \textsl{i.e.}, it gives a larger value when $x \rightarrow
0$ and approaches to $0$ when $x \rightarrow 1$.

Fig.~2 gives the results of $\Lambda$. As is shown,
for ${\Delta}u_{\scriptscriptstyle{V}}(x)/u_{\scriptscriptstyle{V}}(x)$,
SU(3)-extended BBS model gives much smaller prediction than the other
models, while for
${\Delta}s_{\scriptscriptstyle{V}}(x)/s_{\scriptscriptstyle{V}}(x)$
and $u_{\scriptscriptstyle{V}}(x)/s_{\scriptscriptstyle{V}}(x)$,
predictions are close to each other. Such result means
that our extension of BBS model through SU(3) flavor symmetry is
reasonable in the studies of baryon structures.

The comparisons for other baryons are also obtained. In Figs.~3 and
4, we make comparisons for $\Sigma^+$ and $\Sigma^0$. We can see that
in SU(3)-extended BBS model, $\Delta u_{\scriptscriptstyle{V}}(x) / u_{\scriptscriptstyle{V}}(x)$
approaches to $0.7 \sim 0.8$ when $x \rightarrow 1$, allowing
partly negative polarization of u quark, contrary to the result of 1 from the other three models.
Then as for $\Delta
s_{\scriptscriptstyle{V}}(x) / s_{\scriptscriptstyle{V}}(x)$,
SU(3)-extended BBS model and quark-diquark model give results around
$-0.5$, against the positive value given by pQCD and brief pQCD. And as for $
s_{\scriptscriptstyle{V}}(x) / u_{\scriptscriptstyle{V}}(x)$, pQCD, brief pQCD
and SU(3)-extended BBS models give a similarly positive value, while
quark-diquark model predicts zero.

Fig.~5 presents results of $\Xi^-$. The behaviors for $\Delta
s_{\scriptscriptstyle{V}}(x) / s_{\scriptscriptstyle{V}}(x)$ and
$\Delta d_{\scriptscriptstyle{V}}(x) / d_{\scriptscriptstyle{V}}(x)$
are very similar to those of $\Delta u_{\scriptscriptstyle{V}}(x) /
u_{\scriptscriptstyle{V}}(x)$ and $\Delta
s_{\scriptscriptstyle{V}}(x) / s_{\scriptscriptstyle{V}}(x)$ in Fig.~3, respectively.
And as for $d_{\scriptscriptstyle{V}}(x) / s_{\scriptscriptstyle{V}}(x)$,
it is shown that quark-diquark model get a slightly
different behavior from that of the other models.

The results of $n$, $\Sigma^{-}$ and $\Xi^{0}$ are not presented
here, because they can be easily obtained from $p$, $\Sigma^{+}$ and
$\Xi^{-}$ respectively, according to $u$-$d$ isospin symmetry. For
examples, $u^n(x) = d^p(x)$ and $d^{\Sigma^{-}}(x) =
u^{\Sigma^{+}}(x)$.

One can also use other phenomenological parameterizations to study
octet-baryon PDFs from SU(3) relations, and in Ref.~\cite{su3} two
sets of octet-baryon PDFs were obtained from polarized PDFs
parameterizations of the proton.

Besides, our calculation of quark fragmentation functions also gives
predictive results. According to our BBS-based model and the
Gribov-Lipatov relation, we get $D_{u}^{p}(z)/D_{d}^{p}(z) \to
6.252$ and $D_{s}^{\Lambda}(z)/D_{u}^{\Lambda}(z) \to 1.648$ when $z
\to 1$. But in the statistical approach in Ref.~\cite{ff2003}, the
calculation gives $D_{u}^{p}(z)/D_{d}^{p}(z) \to 1$ and
$D_{s}^{\Lambda}(z)/D_{u}^{\Lambda}(z) \to 25.217$. We can see that
the difference is very large both for the proton and $\Lambda$. We
can also compare with AKK (Albino-Kniehl-Kramer) phenomenological
parameterization~\cite{akk} of the proton and $\Lambda$
fragmentation functions, and the calculation results are both 0
corresponding to the above two quantities. Thus further
discriminations between different predictions are needed by
experiments.

Finally, we would like to estimate the effect introduced by SU(3)
breaking and Melosh-Wigner rotation~\cite{ma91_93} to the
SU(3)-extended BBS model. In the quark-diquark model, this effect
comes from different masses of u, d and s quarks, and the correction
is roughly estimated to be of about 10\%. We therefore expect that
such effect may bring about the same size correction in the
SU(3)-extended BBS model. It means that our results are
qualitatively reasonable, with some freedom to improve
quantitatively when more data are available.
\section{Conclusion}\label{seccon}

In this paper, we calculate parton distribution functions (PDFs) of
octet baryons through SU(3) flavor symmetry, starting from PDFs of
the proton in the BBS statistical model. We make comparisons among
different models, and find that our combination of the BBS
statistical model and SU(3) flavor symmetry is an optional way to
explore hadron structures. We also calculate the fragmentation
function (FF) ratios, $D_{u}^{p}(z)/D_{d}^{p}(z)$ and
$D_{u}^{\Lambda}(z)/D_{s}^{\Lambda}(z)$, at the $z \to 1$ region,
based on the phenomenological Gribov-Lipatov relation. The results
are compared with those from predictions of other models. We find
our results are reasonable and the method can be extended to the
studies of baryon structures.

\section{Acknowledgements}\label{secac}

The work is partially supported by National Natural Science
Foundation of China (11005018, 11021092, 10975003, 11035003).

\begin{largetable}
\caption{PDFs of octet baryons in the quark-diquark model. The unit
of $m_{\scriptscriptstyle{D}}$ ($D=V$ or $S$) is MeV, and $m_{q}
=330~\textrm{MeV}$ for $u$, $d$ quarks, $m_{q} =480$~MeV for $s$
quark.} \label{1}
\begin{center}
\begin{tabular}  {cllcc}
\hline \hline%
& $q$  &  ${\Delta}q$    &    $m_{\scriptscriptstyle{V}}$  &$m_{\scriptscriptstyle{S}}$         \\
\hline%
$p$ & $u=a_{\scriptscriptstyle{V}}/6+a_{\scriptscriptstyle{S}}/2$ & ${\Delta}u=-\tilde{a}_{\scriptscriptstyle{V}}/18+a_{\scriptscriptstyle{S}}/2$   &800 &600\\
    & $d=a_{\scriptscriptstyle{V}}/3$                                      & ${\Delta}d=-\tilde{a}_{\scriptscriptstyle{V}}/9$                                         &800&600 \\
\hline%
$\Sigma^{+}$   & $u=a_{\scriptscriptstyle{V}}/6+a_{\scriptscriptstyle{S}}/2$&${\Delta}u=-\tilde{a}_{\scriptscriptstyle{V}}/18+\tilde{a}_{\scriptscriptstyle{S}}/2$   &950 &750 \\
                    & $s=a_{\scriptscriptstyle{V}}/3$ &${\Delta}s=-\tilde{a}_{\scriptscriptstyle{V}}/9$   &800    &600 \\
\hline
$\Sigma^{0}$& $u=a_{\scriptscriptstyle{V}}/12+a_{\scriptscriptstyle{S}}/4$   &${\Delta}u=-\tilde{a}_{\scriptscriptstyle{V}}/36+\tilde{a}_{\scriptscriptstyle{S}}/4$    &950 &750 \\
                 & $d=a_{\scriptscriptstyle{V}}/12+a_{\scriptscriptstyle{S}}/4$ &${\Delta}d=-\tilde{a}_{\scriptscriptstyle{V}}/36+\tilde{a}_{\scriptscriptstyle{S}}/4$    &950 &750  \\
                 & $s=a_{\scriptscriptstyle{V}}/3$ &${\Delta}s=-\tilde{a}_{\scriptscriptstyle{V}}/9$                                          &800    &600 \\
\hline
$\Lambda^{0}$    & $u=a_{\scriptscriptstyle{V}}/4+a_{\scriptscriptstyle{S}}/12$   &${\Delta}u=-\tilde{a}_{\scriptscriptstyle{V}}/12+\tilde{a}_{\scriptscriptstyle{S}}/12$    &950 &750 \\
& $d=a_{\scriptscriptstyle{V}}/4+a_{\scriptscriptstyle{S}}/12$   &${\Delta}d=-\tilde{a}_{\scriptscriptstyle{V}}/12+\tilde{a}_{\scriptscriptstyle{S}}/12$    &950 &750 \\
                 & $s=a_{\scriptscriptstyle{S}}/3$ &${\Delta}s=\tilde{a}_{\scriptscriptstyle{S}}/3$      &800 &600\\
\hline
$\Xi^{-}$     & $d=a_{\scriptscriptstyle{V}}/3$    &${\Delta}d=-\tilde{a}_{\scriptscriptstyle{V}}/9$ &1100 &900\\
                   & $s=a_{\scriptscriptstyle{V}}/6+a_{\scriptscriptstyle{S}}/2$  &${\Delta}s=-\tilde{a}_{\scriptscriptstyle{V}}/18+\tilde{a}_{\scriptscriptstyle{S}}/2$  &950 &750\\
\hline
\end{tabular}
\end{center}
\end{largetable}

\begin{largetable}
\caption{Parameters for PDFs of octet baryons in pQCD.} \label{2}
\begin{center}
\begin{tabular}{ccccccccccc}
\hline\hline
& $q_{1}$  &$q_{2}$ & $\tilde{A}_{q_{1}}$ &$\tilde{B}_{q_{1}}$ & $\tilde{C}_{q_{1}}$ & $\tilde{D}_{q_{1}}$ & $\tilde{A}_{q_{2}}$ &$\tilde{B}_{q_{2}}$ & $\tilde{C}_{q_{2}} $ & $\tilde{D}_{q_{2}}$\\
\hline
$p$&$u$ &$d$ &5.0 &-3.61 &3.0 &-2.40 &1.0 &-0.74 &2.0 &-1.27\\
$\Sigma^{+}$ &$u$ &$s$ &5.0 &-3.61 &3.0 &-2.40 &1.0 &-0.74 &2.0 &-1.27\\
$\Sigma^{0}$&$u(d)$ &$s$  &2.0 &-1.34 &2.0 &-1.67 &0.8 &-0.54 &2.0 &-1.27\\
$\Lambda^{0}$  &$s$ &$u(d)$ &2.0   &-1.2  &2.0 &-1.8 &1.0 &-0.6 &2.0 &-1.40\\
$\Xi^{-}$ &$s$  &$d$   &5.0 &-3.61 &3.0 &-2.40 &1.0 &-0.74 &2.0 &-1.27\\
\hline
\end{tabular}
\end{center}
\end{largetable}

\begin{largetable}
\caption{Parameters for PDFs of octet baryons in the brief
pQCD.}\label{pqcd2}
\begin{center}
\begin{tabular}{cccccccccc}
\hline \hline
                 & $q_{1}$  &$q_{2}$  &$R_{A}$ &${\Delta}q_{1}$ &${\Delta}q_{2}$ & $\tilde{A}_{q_{1}}$   &$\tilde{C}_{q_{1}}$     &  $\tilde{A}_{q_{2}}$    &$\tilde{C}_{q_{2}} $           \\
\hline
$p$ &$u$ &$d$ &5 &0.75  &-0.45 &1.375 &0.625 &0.275 &0.725\\
$\Sigma^{+}$&$u$  &$s$  &5 &0.75  &-0.45 &1.375 &0.625 &0.275 &0.725\\
$\Sigma^{0}$ &$u(d)$ &$s$  &$5/2$ &0.375 &-0.45 &0.6875 &0.3125 &0.275 &0.725\\
$\Lambda^{0}$ &$s$ &$u(d)$ &2  &0.65 &-0.175 &0.825 &0.175 &0.4125 &0.5875\\
$\Xi^{-}$ &$s$  &$d$  &5 &0.75 &-0.45 &1.375 &0.625 &0.275 &0.725\\
\hline
\end{tabular}
\end{center}
\end{largetable}

\begin{largetable}
\caption{PDFs of octet baryons from SU(3) flavor
symmetry.}\label{symmetry}
\begin{center}
\begin{tabular}{ccccccc}\hline \hline
                & $u_{v}^{B}$  &$d_{v}^{B}$&  $s_{v}^{B}$& $\bar{u}^{B}$&         $\bar{d}^{B}$&           $\bar{s}^{B}$ \\
\hline
   p                  & $u_{v}$      &  $d_{v}$               & $s-\bar{s}$    &$\bar{u}$   &$\bar{d}$                  & $\bar{s}$              \\
$\Sigma^{+}$      &$u_{v}$         &  $s-\bar{s}$           &$d_{v}$         &$\bar{u}$    &$\bar{s}$                &$\bar{d}$ \\
$\Sigma^{0}$& $\left(u_{v}+s-\bar{s}\right)/2$ & $\left(u_{v}+s-\bar{s}\right)/2$ &$d_{v}$&$\left(\bar{u}+\bar{s}\right)/2$&$\left(\bar{u}+\bar{s}\right)/2$&$\bar{d}$\\
$\Lambda^{0}$&$\left(u_{v}+4d_{v}\right)/6$&$\left(u_{v}+4d_{v}\right)/6$&$\left(2u_{v}-d_{v}\right)/3$  &$\left(\bar{u}+\bar{s}\right)/2$&$\left(\bar{u}+\bar{s}\right)/2$&$\bar{d}$\\
$\Xi^{-}$     &$s-\bar{s}$      &$d_{v}$      &$u_{v}$       &$\bar{s}$      &$\bar{d}$       &$\bar{u}$\\
\hline
\end{tabular}
\end{center}
\end{largetable}

\begin{figure*}[htbp]
\centering
\includegraphics[width=2in]{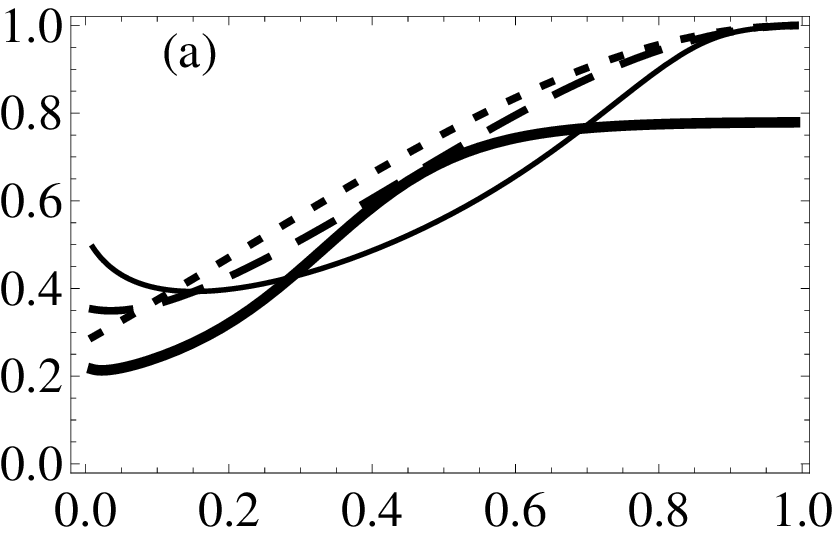}
\includegraphics[width=2.1in]{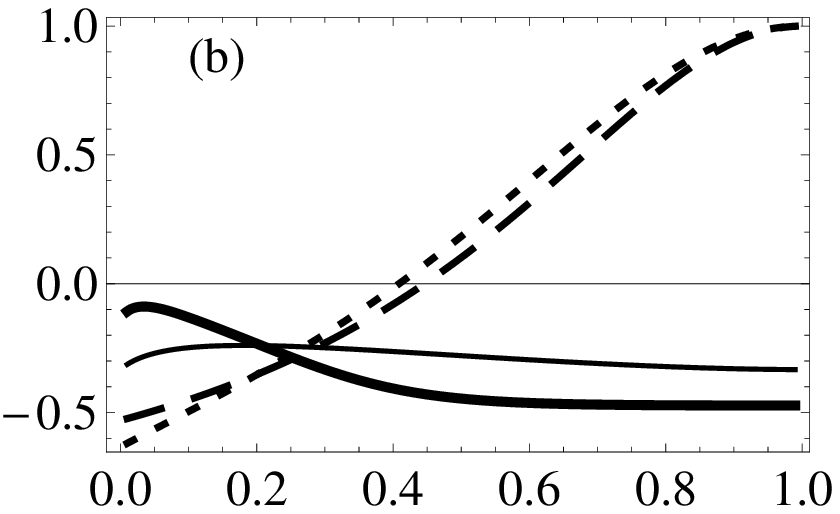}
\includegraphics[width=2in]{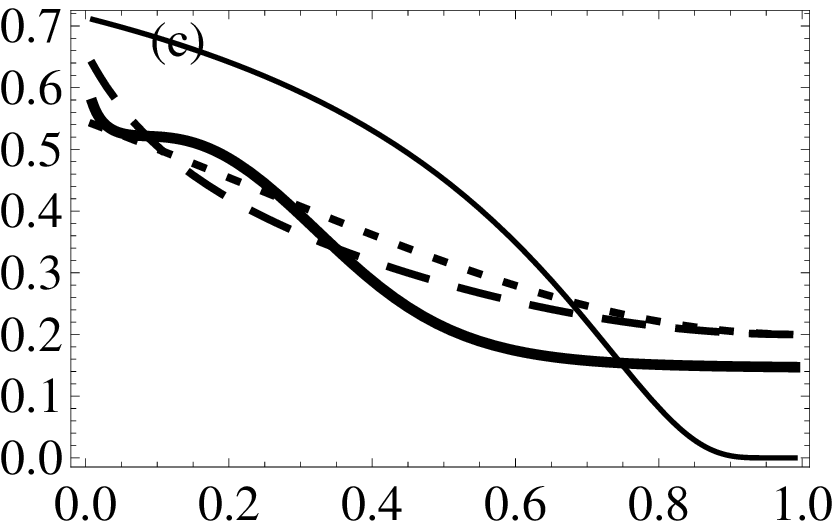}
\caption{(a) ${\Delta}u(x)/u(x)$, (b) ${\Delta}d(x)/d(x)$, (c)
$d(x)/u(x)$ for the proton. The full, dotted, dashed, and thick full
lines are results for quark-diquark model, brief pQCD model, pQCD
model, and BBS model, respectively.} \label{fig}
\end{figure*}
\begin{figure*}[htbp]
\centering
\includegraphics[width=2.1in]{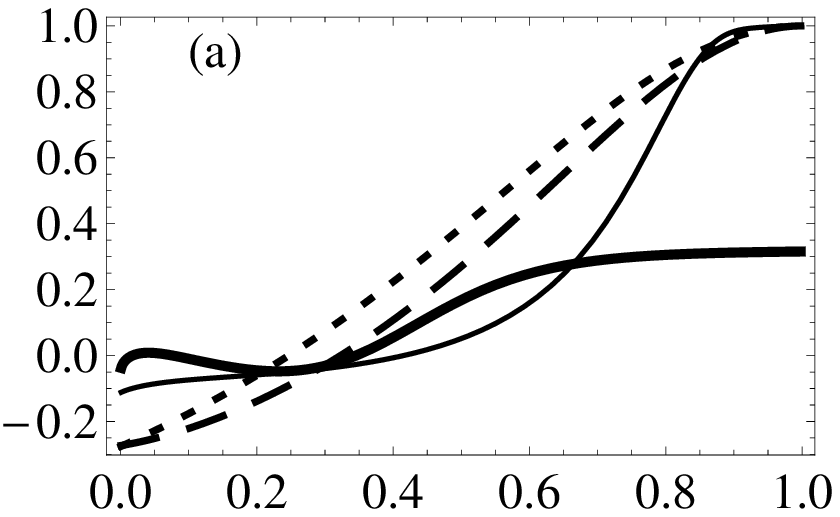}
\includegraphics[width=2in]{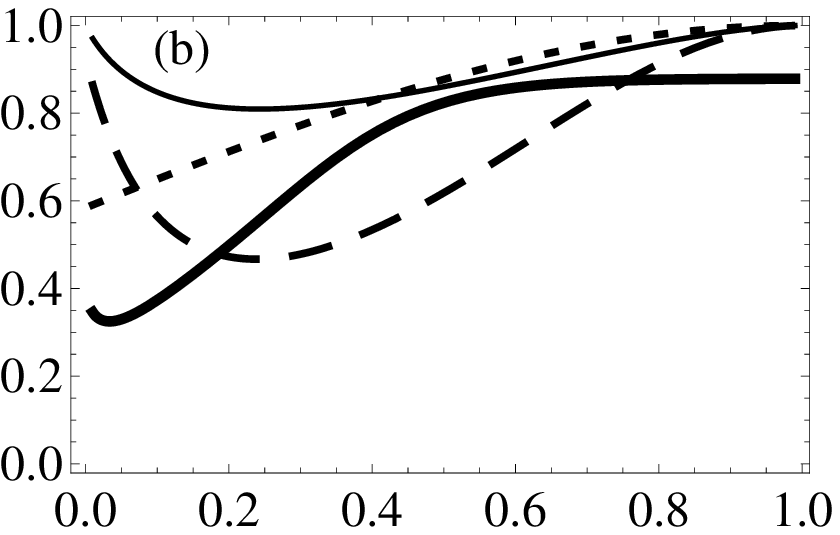}
\includegraphics[width=2in]{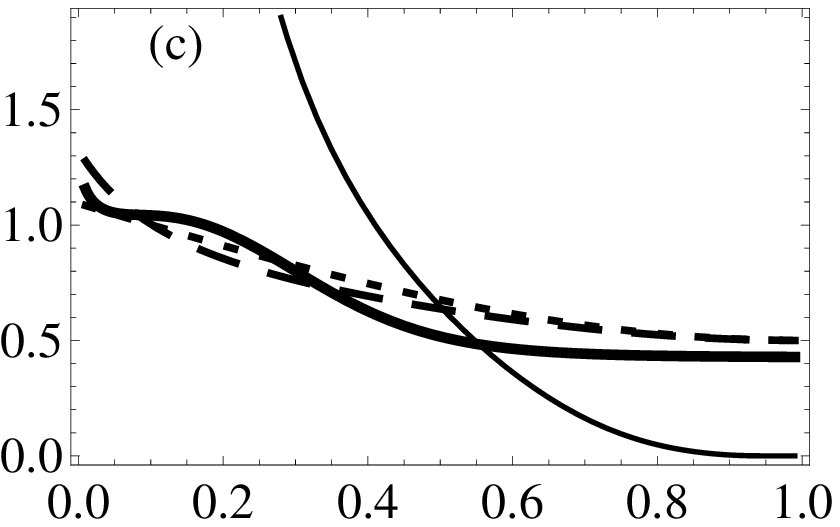}
\caption{(a) ${\Delta}u(x)/u(x)$, (b) ${\Delta}s(x)/s(x)$, (c)
$u(x)/s(x)$ for $\Lambda$. The full, dotted, dashed, and thick full
lines are results for quark-diquark model, brief pQCD model, pQCD
model, and SU(3)-extended BBS model, respectively.} \label{fig}
\end{figure*}
\begin{figure*}[htbp]
\centering
\includegraphics[width=2in]{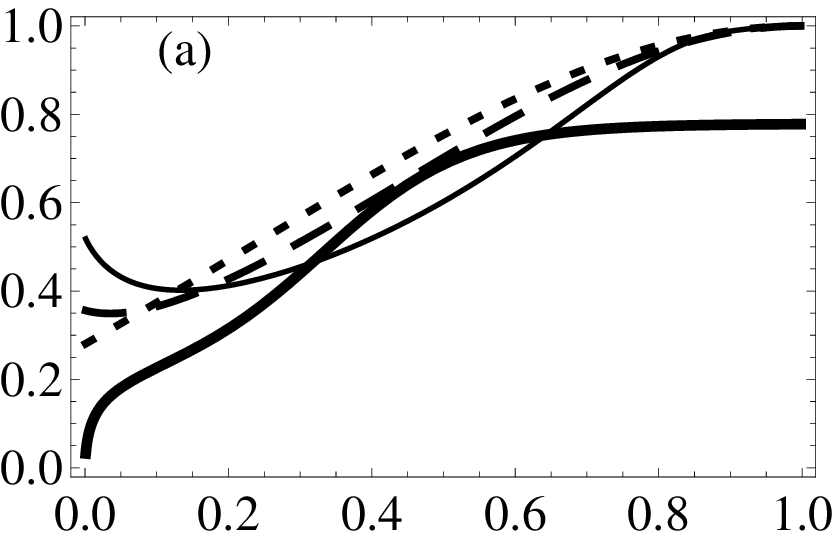}
\includegraphics[width=2.1in]{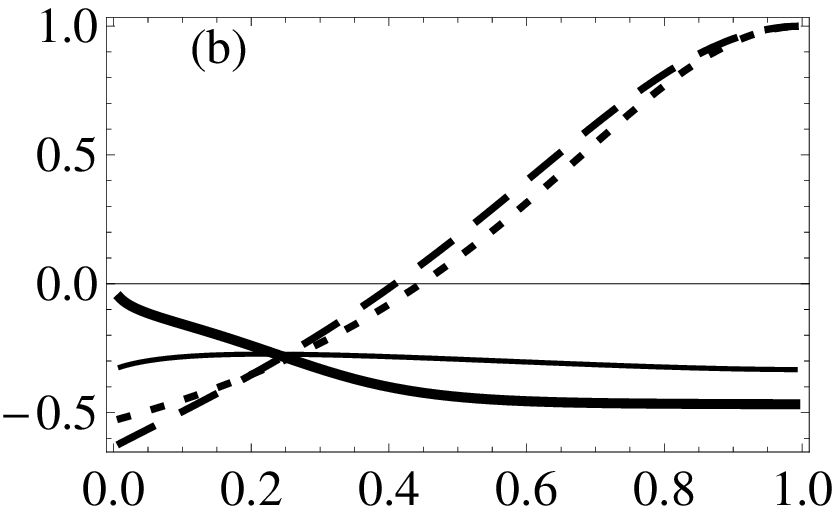}
\includegraphics[width=2in]{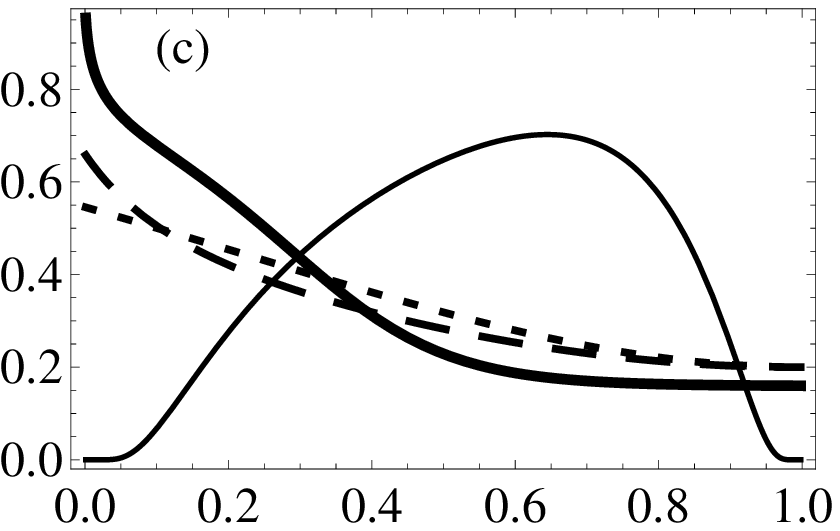}
\caption{(a) ${\Delta}u(x)/u(x)$, (b) ${\Delta}s(x)/s(x)$, (c)
$s(x)/u(x)$ for $\Sigma^{+}$. Models are identical to those in
Fig.~$2$.} \label{fig}
\end{figure*}
\begin{figure*}[htbp]
\centering
\includegraphics[width=2in]{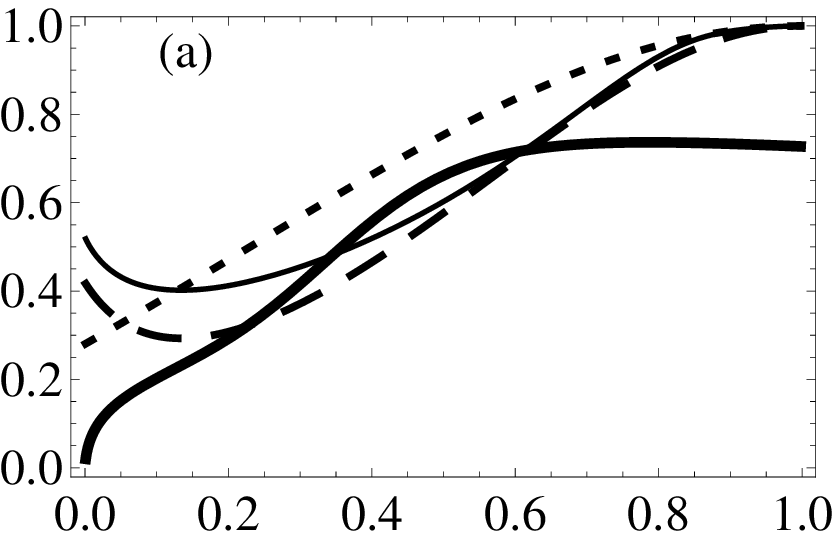}
\includegraphics[width=2.1in]{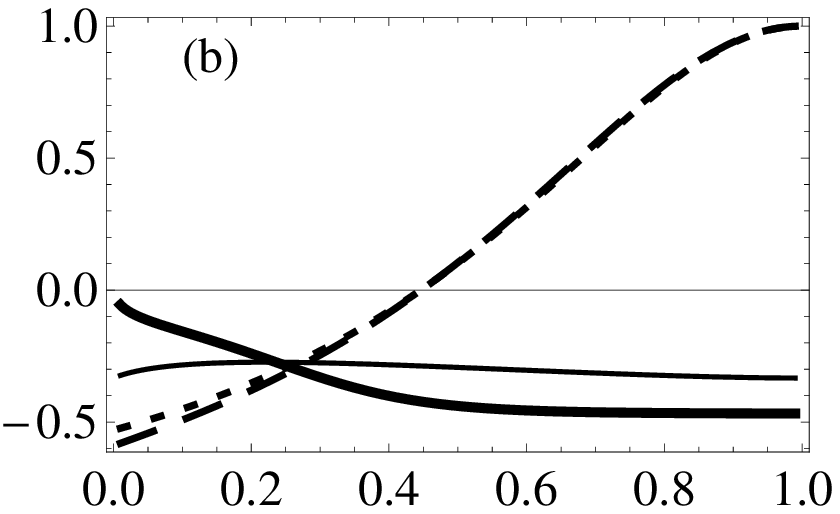}
\includegraphics[width=2in]{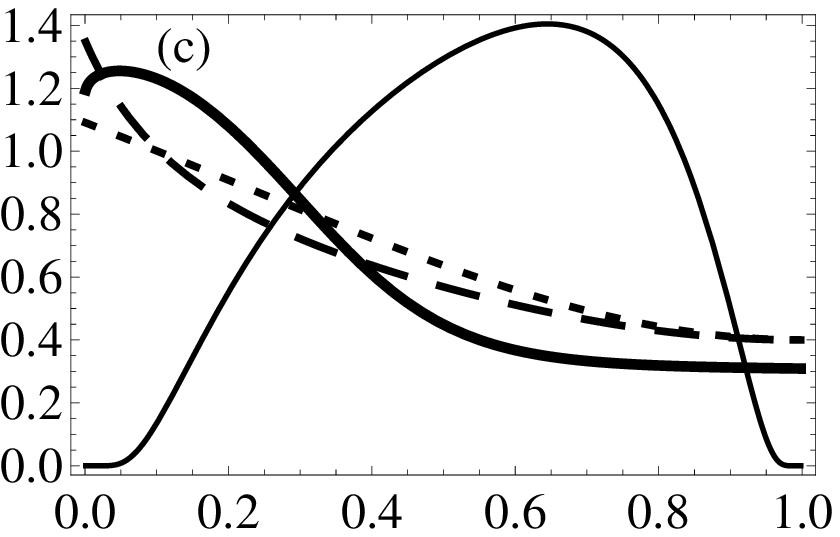}
\caption{(a) ${\Delta}u(x)/u(x)$, (b) ${\Delta}s(x)/s(x)$, (c)
$s(x)/u(x)$ for $\Sigma^{0}$. Models are identical to those in
Fig.~$2$.} \label{fig}
\end{figure*}
\begin{figure*}[htbp]
\centering
\includegraphics[width=2in]{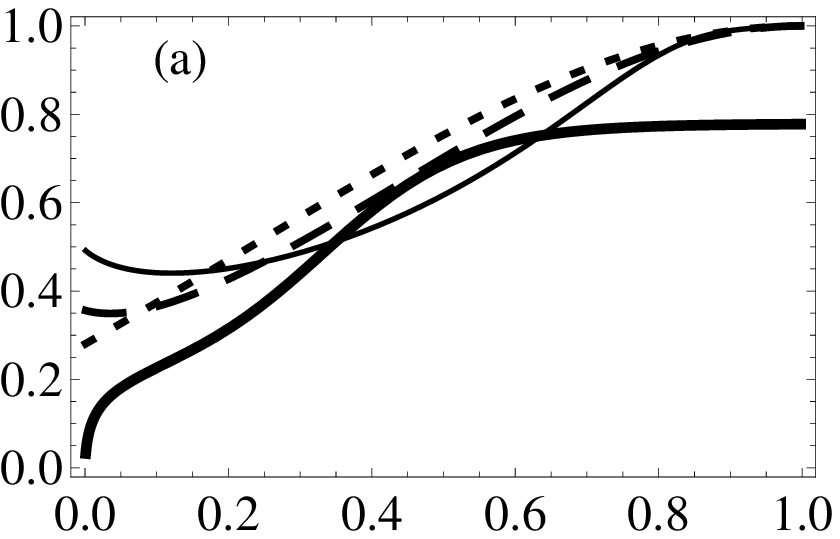}
\includegraphics[width=2.1in]{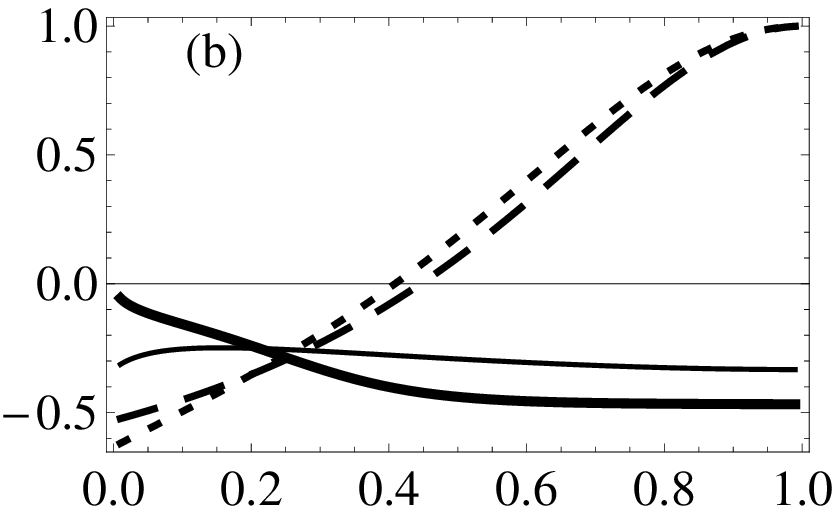}
\includegraphics[width=2in]{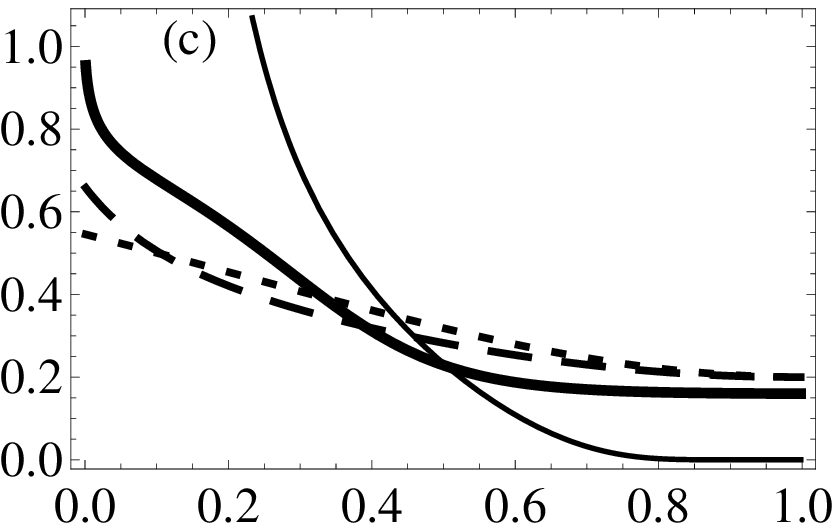}
\caption{(a) ${\Delta}s(x)/s(x)$, (b) ${\Delta}d(x)/d(x)$, (c)
$d(x)/s(x)$ for $\Xi^{-}$. Models are identical to those in
Fig.~$2$.}
\end{figure*}

\end{document}